# MATHEMATICAL TUTORIALS
# IN INTRODUCTORY PHYSICS

## Richard N. Steinberg, Michael C. Wittmann, and Edward F. Redish
*Department of Physics, University of Maryland, College Park, MD 20742-4111*

Students in introductory calculus-based physics not only have difficulty understanding the fundamental physical concepts, they often have difficulty relating those concepts to the mathematics they have learned in math courses. This produces a barrier to their robust use of concepts in complex problem solving. As a part of the *Activity-Based Physics* project, we are carrying out research on these difficulties and are developing instructional materials in the tutorial framework developed at the University of Washington by Lillian C. McDermott and her collaborators. In this paper, we present a discussion of student difficulties and the development of a mathematical tutorial on the subject of pulses moving on strings.

## INTRODUCTION

Students in introductory calculus-based physics are well-known to have difficulty in developing a good understanding of fundamental physical concepts.[1] But our observations show that even when instruction is modified in a way that improves students' understanding of concepts, they often have difficulty in applying those concepts in the context of complex problem solving. One reason for this seems to be that students often have difficulty in linking the relevant mathematical equations and methods to those concepts.[2]

As a part of the *Activity-Based Physics* project,[3] we are developing a series of instructional materials in introductory physics to address these problems. We call these materials *Mathematical Tutorials*. These tutorials are intended as a supplement to *Tutorials in Introductory Physics*, a research and curriculum development project by Lillian C. McDermott and the Physics Education Group at the University of Washington.[4,5,6] (The University of Maryland is a test site for this NSF project.[7]) Tutorials are a set of instructional materials intended to supplement the lectures and textbook of a standard physics course. They are developed in a tight cycle with instruction and physics education research that results in materials that deal effectively with specific student difficulties.

The context for both the research and curriculum development described in this paper is the introductory calculus-based physics course at the University of Maryland. This is a three semester sequence with three hours of lecture and one hour of either recitations or tutorials each week. Calculus II is a co-requisite for the first course in the sequence, but many students are even further along in their math courses and many of them are doing quite well in them.

A sample set of instructional materials is included in the appendices. In the rest of this paper, we give an overview of tutorials and describe the development of a mathematical tutorial.

# TUTORIALS

The components of the tutorials are a 10-minute ungraded pretest, a 50-minute tutorial session, and a tutorial homework assignment. The pretest asks qualitative conceptual questions about the subject to be covered in the subsequent tutorial. Its purpose is to help both the student and the instructor recognize gaps in the student's understanding. It is usually given after lectures and readings on the subject, but before the tutorial. In the tutorial sessions, students work in groups of three or four and answer questions on a worksheet that guides them through building qualitative reasoning on fundamental concepts. Teaching assistants serve as facilitators, asking questions that help the students work through difficulties in their own thinking. On the homework assignment, students apply the ideas covered in the tutorial session and continue to reason qualitatively and explain their reasoning. For a detailed description of tutorials and their development at the University of Washington, see refs. 5 and 6. For a description of the implementation and success of tutorials at the University of Maryland, see ref. 7.

It has been the experience of both the University of Washington group and our group that it is crucial to support the tutorial project by conducting a weekly training session. All tutorial instructors complete the same pretest and tutorials as the students. They read and discuss student responses to the pretest. Issues of content, pedagogy, and instructional strategies are all considered in the context of actually doing a tutorial. Both groups have also found that it is essential to ask a question emphasizing material from tutorials on each examination. For the students, this highlights the importance of the kinds of skills developed in the tutorials. For the instructors, it provides continuing feedback regarding student understanding of the subject matter.

# DEVELOPMENT OF A MATHEMATICAL TUTORIAL

It has been shown that students at many levels have difficulties understanding fundamental concepts of mathematics.[8,9] In our investigations of student difficulties with mathematics in introductory physics, we have found that there are a number of fundamental ideas covered in math classes that many students are consistently unable to apply in physics. We are developing a series of tutorials based on this research that focus on these difficulties. As an example, we present the research and curriculum development in the context of mechanical waves.

## Student Difficulties With Mathematics
## When Learning Mechanical Waves

The subject of mechanical waves is one of considerable importance in the introductory calculus-based physics course. It is the student's first introduction to wave theory, a subject which will recur in many contexts including physical optics, acoustics, and quantum mechanics. Furthermore, it is the student's first experience with trying to interpret the mathematics of a function of two independent variables, and this leads to a number of possibilities for confusion and misinterpretation.

We have obtained information on student difficulties with mathematics through a variety of methods. Interaction with students in classroom and informal settings as well as analysis of examination problems provided preliminary insights. This was followed by an iterative development, administration, and analysis of pretests, examination questions, and videotaped individual demonstration interviews. The interviews entailed asking students to make predictions and give explanations about simple physical situations shown to them. They ranged in length from 40 minutes to one hour. All interviews described in this paper were conducted after instruction in the topics included in the interview. The students interviewed were volunteers generally performing above the mean of the class. Although we focus on mechanical waves, the results described in this section are indicative of student difficulties we have seen in many domains.

After traditional instruction on mechanical waves, we administered the question shown in Fig. 1 as both an interview protocol ($N=9$) and as part of a pretest ($N=57$). The interview students were from a different lecture class than the students who took the pretest.

What we considered to be a correct response was to show the pulse displaced an amount $x_0$ and the amplitude unchanged, as shown in Fig. 2a. This answer was given by 44% of the students who were interviewed and 56% of the students who took the pretest. Most of the rest of the students (56% of the interview-students and 35% of the pretest-students) drew a pulse displaced an amount $x_0$ but the amplitude decreased, as shown in Fig. 2b. On the surface this appears to be a reasonable response in that it is consistent with what would actually happen as a result of the physical phenomena (not mentioned in the problem) of friction with the imbedding medium and internal dissipation. However the explanations given by students suggest that they are not adding to the physics of the problem but are misinterpreting the

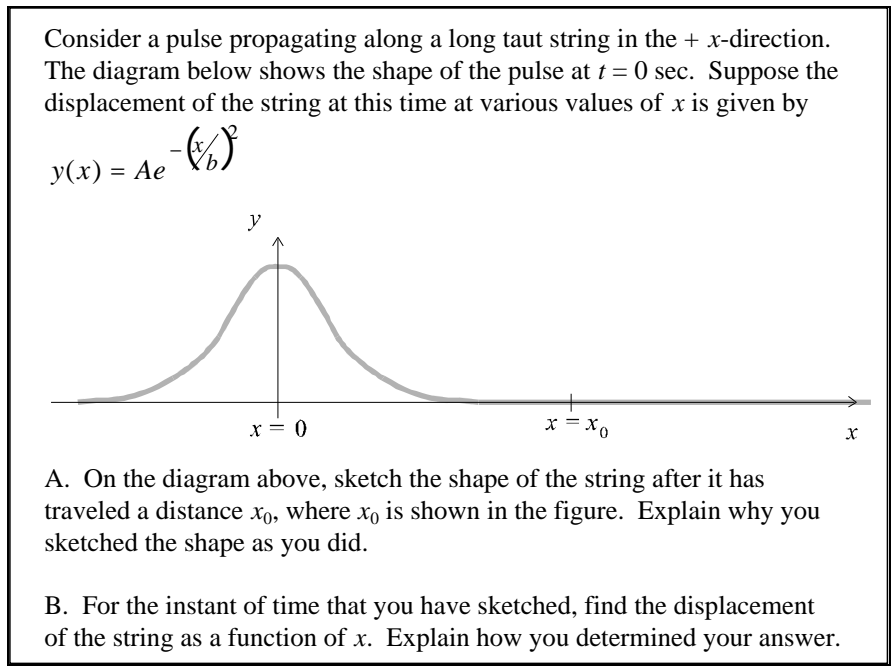

Consider a pulse propagating along a long taut string in the $+\,x$-direction. The diagram below shows the shape of the pulse at $t = 0$ sec. Suppose the displacement of the string at this time at various values of $x$ is given by

$$y(x) = A e^{-\left(x/b\right)^2}$$

A. On the diagram above, sketch the shape of the string after it has traveled a distance $x_0$, where $x_0$ is shown in the figure. Explain why you sketched the shape as you did.

B. For the instant of time that you have sketched, find the displacement of the string as a function of $x$. Explain how you determined your answer.

**FIGURE 1.** Problem used to probe student difficulties with mathematics in the context of mechanical waves.

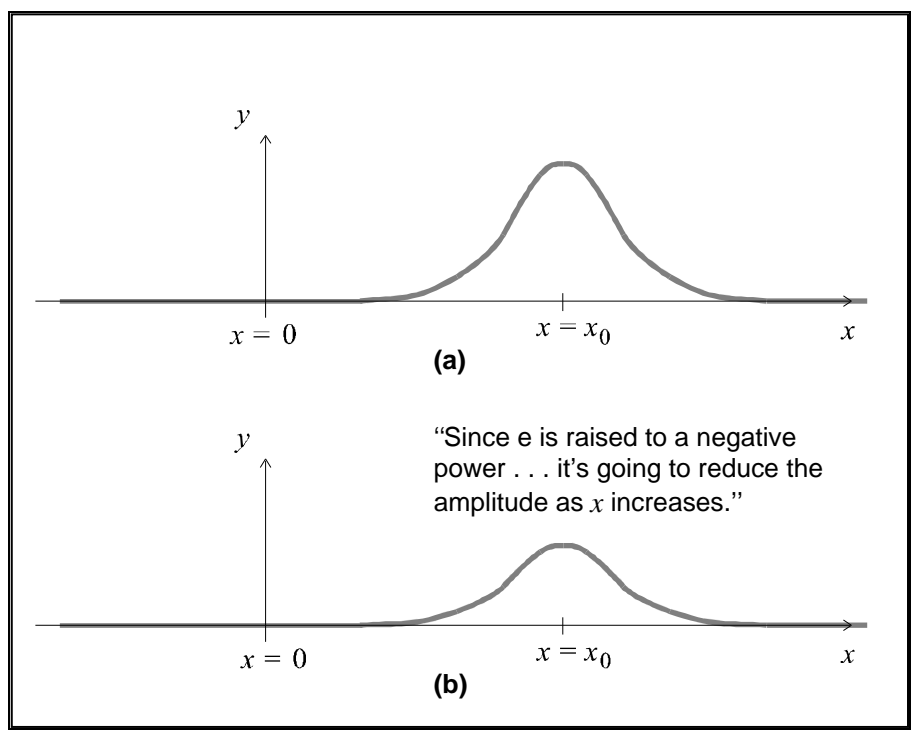

**FIGURE 2.** (a) A correct sketch of the shape of the pulse at a later time, showing the amplitude unchanged, (b) An apparently correct sketch of the shape of the pulse show-ing the amplitude decreased - but typically accompanied by incorrect reasoning.

mathematics. All of the interview students and many of the pretest students cited the equation describing the shape of the string at $t=0$ as the reason for the decrease in the amplitude.[10] As one interviewed student said:

> "Since e is raised to a negative power . . . it's going to reduce the amplitude as $x$ increases."

Unfortunately, the exponential given in this problem represents a decrease in $y$ in space (at $t=0$) not time. These students are failing to recognize that $x$ corresponds to a variable which maps one of the dimensions of the problem, not the location of the peak of the pulse. Without this recognition, a functional understanding of the mathematics of this problem seems unlikely.

Part B of this problem asked about the mathematical form of the string at a later time. We considered any answer that replaced $x$ with $x$-$x_0$ to be correct. However, none of the interview students and fewer than 10% of the pretest students answered this way (see Table 1). The most common incorrect response was to simply substitute $x_0$ for $x$ in the given equation. We refer to this response as a "non-function."[11] It was given by 67% of the interview students and 44% of the pretest students. All students drew a string with different values of $y$ at different values of $x$, yet many of them wrote an equation for that shape with no $x$ dependence. Again this answer is suggestive of misinterpreting $x$ as described above. There were other students who wrote a sinusoidal dependence for $y$, again in conflict with what they drew for the shape of the string.[12] Many students seemed to be answering the mathematical part of this problem independently of the way they were answering the physical part.

**Table 1.** Student responses to give an equation describing the shape of the pulse

| | example(s) | % of interview respondents (N=9) | % of pretest respondents (N=57) |
|---|---|---|---|
| **correct response** | $y(x) = Ae^{-\frac{x-x_0}{b}^2}$ | 0% | 7% |
| **"non-function"** | $y(x) = Ae^{-x/b^2}$ ; $y(x_0) = Ae^{-x/b^2}$ | 67% | 44% |
| **sinusoidal** | $y(x) = A\sin(kx - wt)$ | 22% | 2% |
| **other** | $x = -b \ln \frac{y}{A}$ ; $\frac{dy}{dx} = -\frac{2x}{a^2} Ae^{-(x/a)^2}$ | 11% | 47% |

The difficulties described in this section include students failing to recognize the relationship between the physical situation and the associated equation, failing to understand the meaning of a function, and failing to treat a coordinate axis as a mapping of a dimension. Note that these results occurred after instruction which both included lectures addressing the problem and readings of a standard text. We have seen similar difficulties in other domains and have been developing mathematical tutorials to address these difficulties as part of the *Activities-Based Physics* project.

## Overview of a Tutorial: *Mathematical description of pulses*

The tutorial process begins with students working independently on pretests. The pretest for this tutorial is based on the problem shown in Fig. 1 and is given in its entirety in Appendix A. In taking the pretest, students are forced to think through the problem on their own, commit to an answer, and articulate that answer in writing. As suggested in the previous section, the problem is both challenging and relevant.

In tutorial, groups of three or four students work through guided worksheets. The worksheet for this tutorial is included in Appendix B. Students begin by considering the mathematical form of a pulse at $t=0$. In order to minimize the confusion related to the exponential, we begin this tutorial with the equation:

$$y(x) = \frac{50cm}{\left(\frac{x}{b}\right)^2 + 1}$$

(1)

where $b=20$ $cm$. In order to help students develop a functional understanding of a function, they explicitly graph the shape of the string based on the equation representing its shape. They then sketch the shape the string would have after the pulse traveled some distance without dissipation, and are guided into constructing the mathematical form. In this way, students not only construct the shape of the string from an equation, they construct an equation from the shape of a string. After considering the functional form of the pulse at two different times, the students are given the opportunity to construct a single equation that describes the pulse as a function of both position and time. The key here is that it is the students that are constructing this equation based on their own work and on consideration of a specific physical system.

In the next part of the tutorial, students apply and interpret the ideas that they developed by considering the motion of a tagged part of the string. Here they extract useful information about the motion of the tag by interpreting the mathematics of the problem.

In the third part of the tutorial, students consider a pulse of a slightly different shape propagating on a string. In particular, they consider a pulse represented at $t=0$ by the same equation considered in the pretest:

$$y(x) = Ae^{-\left(x/b\right)^2}$$

(2)

Students again are asked to construct an equation that describes the displacement of the string as a function of position and time. This time, students are not guided to this answer. Instead, students are forced to generalize their results from earlier in the tutorial and, when appropriate, resolve the conflict with their answers on the pretest.

In the previous section, we described student difficulties relating the physical situation and the corresponding equation. Because of these difficulties, in the last part of the tutorial students use video software to mathematically model the shape of an actual pulse. As part of this, they explore the physical significance of the parameters $A$ and $b$ in equation (2).

The homework which accompanies this tutorial is given in Appendix C. Students apply the ideas covered in the tutorial while considering pulses of different shapes moving in different directions. They consider the shape of the pulses at different times physically and mathematically. They also consider the motion of the tagged part of the string in different situations.

# MATHEMATICAL TUTORIALS IN INTRODUCTORY PHYSICS

In addition to *Mathematical Description of Pulses*, we are developing several other mathematical tutorials throughout the entire three semester sequence. This gives students repeated opportunities to understand fundamental mathematical issues, such as those described in this paper. These tutorials have several common features. They address difficulties students have with understanding and applying mathematical ideas when learning introductory physics. They are developed within the tutorial framework and philosophy described in this paper and refs. 4-6. They are all based on the results of physics education research. The mathematical tutorials do not merely hone student skills with the mathematics, they do so in the context of students interpreting the fundamental physics concepts which they are currently learning. In addition, the mathematical tutorials enable us to include components in the other tutorials which build on ideas such as those described in this paper.

# CONCLUSION

In this paper, we have described how we investigate student difficulties with mathematics and use the results of this research as a guide to curriculum development. In the tutorial format that we employ, the curriculum is developed with awareness of student difficulties. The activities are designed so that the students are engaged and responsible for their own construction of ideas. It is our belief that learning is enhanced in this type of setting.

We are currently using the curriculum described in this paper in the introductory calculus-based physics course at the University of Maryland. In addition to carrying out preliminary studies on the effectiveness of our materials, we are continuing to investigate the nature of student difficulties with the subject matter, including in courses using tutorials. The results of this continued research feeds back into the curriculum development as part of an iterative process of research, curriculum development, and instruction.

# ACKNOWLEDGMENTS

The authors would like to thank Mel Sabella and Jeff Saul for their contributions to the research described in this paper. This work is suported in part by NSF grant DUE-9455561.

# ENDNOTES

[1] See for example, McDermott, L. C., "Millikan Lecture 1990: What we teach and what is learned — Closing the gap," *Am. J. Phys.* **59** 301-315, 1991; Reif, F., "Millikan Lecture 1994: Understanding and teaching important scientific thought processes," *Am. J. Phys.* **63** 17-32, 1995.

[2] See Redish, E. F., Steinberg, R. N., and Saul, J. M., "Student Difficulties with Math in Physics: Giving Meaning to Symbols," AAPT Announcer **26**:2, 70, 1996 and Saul, J. M., Wittmann, M. C., Steinberg, R. N., and Redish, E. F., "Student Difficulties with Math in Physics: Why Can't Students Apply What They Learn in Math Class?" AAPT Announcer **26**:2, 70, 1996. This research has been done as part of the project, "Expectations in University Physics," NSF grant RED-9355849.

[3] *Activity-Based Physics* is a multi-university consortium to develop materials for introductory college physics, sponsored by the National Science Foundation.

[4] McDermott, L. C., Shaffer, P. S., and the Physics Education Group at the University of Washington, *Tutorials in Introductory Physics*, to be published 1997.

[5] A discussion of the tutorials and the role of research in their development can be found in McDermott, L.C., "Bridging the gap between teaching and learning: The role of research," also in this volume. In addition, a sample class on the tutorials was presented at this conference. See McDermott, L.C., Vokos, S., and Shaffer, P. S., "Sample Class on Tutorials in Introductory Physics," in these Proceedings.

[6] For other examples of tutorials and of the research that underlies their development, see, McDermott, L.C., Shaffer, P. S., and Somers, M. D., "Research as a guide for teaching introductory mechanics: An illustration in the context of the Atwood's machine," *Am. J. Phys.* **62**, 46-55, 1994; McDermott, L.C. and Shaffer, P. S., "Research as a guide for curriculum development: an example from introductory electricity, Part I:

Investigation of student understanding," *Am. J. Phys.* **60**, 994-1003, 1992; Shaffer, P. S. and McDermott, L. C., "Research as guide for curriculum development: an example from electricity, Part II: Design of instructional strategies," *Am. J. Phys.* **60**, 1003-1013, 1992.

[7] Redish, E. F., Saul, J. M., and Steinberg, R. N., "On the effectiveness of active-engagement microcomputer-based laboratories," to be published in *Am. J. Phys.*

[8] For examples of work done by the mathematics education research community, see Grouws,D.A., *Handbook of Research on Mathematics Teaching and Learning*, New York: Macmillan Publishing Co., 1992.

[9] There has been some work on student understanding of mathematics in the context of physics, but most is related to interpreting graphs. For examples, see Beichner, R. J., "Testing student interpretation of kinematic graphs," *Am. J. Phys.* **62** 750-762, 1994; McDermott, L.C., Rosenquist, M. L., and van Zee, E. H., "Student difficulties in connecting graphs and physics: Examples from kinematics," *Am. J. Phys.* **55**, 503-515, 1987.

[10] In the interview format, we had the opportunity to obtain an explanation from all of the students. In the pretests, not all of the students give explanations, but those who do cite the exponential as the reason for the decay.

[11] We use this term despite the obvious fact that a constant can be thought of as a function. The students are misinterpreting the meaning of the function involved and replacing it by a value describing one aspect of the function. No student indicated either in writing or in the interviews that they interpreted the string's later shape as a "constant function" (uniform displacement of the string).

[12] A modified version of this question was asked on a midterm examination (to a different class). In this case, 45% of the students gave a sinusoidal answer to mathematically describe the shape of a pulse.